\documentclass{aa}
\usepackage{graphicx}
\usepackage{txfonts}
\usepackage{natbib}
\bibpunct{(}{)}{;}{a}{}{,} 

\begin{document}
\title{K-band spectroscopy of pre-cataclysmic variables\thanks{based on
observations made at ESO-Paranal, proposals 075.D-0012 and 076.D-0538.}}

\author{
C. Tappert\inst{1},
B. T. G\"ansicke\inst{2},
L. Schmidtobreick\inst{3},
R. E. Mennickent\inst{4},
F. P. Navarrete\inst{1}
}

\authorrunning{C. Tappert et al.}

\offprints{C. Tappert}

\institute{
Departamento de Astronom\'{\i}a y Astrof\'{\i}sica, Pontificia Universidad
Cat\'olica, Vicu\~na Mackenna 4860, 782-0436 Macul, Chile\\
\email{ctappert@astro.puc.cl, fpnavarr@puc.cl}
\and
Department of Physics, University of Warwick, Coventry CV4 7AL, UK\\
\email{Boris.Gaensicke@warwick.ac.uk}
\and
European Southern Observatory, Casilla 19001, Santiago 19, Chile\\
\email{lschmidt@eso.org}
\and
Departamento de F\'{\i}sica, Universidad de Concepci\'on, Casilla 160-C,
Concepci\'on, Chile\\
\email{rmennick@astro-udec.cl}
}

\date{Received xxx; accepted xxx}

\abstract
{}
{
There exists now substantial evidence for abundance anomalies in a number of 
cataclysmic variables (CVs), indicating that the photosphere of the secondary
star incorporates thermonuclear processed material. However, the spectral 
energy distribution in CVs is usually dominated by the radiation produced by 
the accretion process, severely hindering an investigation of the stellar 
components. On the other hand, depending on how the secondary star has
acquired such material, the above mentioned abundance anomalies could also be 
present in pre-CVs, i.e.\, detached white/red dwarf binaries that will 
eventually evolve into CVs, but have not yet started mass transfer, and 
therefore allow for an unobstructed view on the secondary star at infrared 
wavelengths.
}
{
We have taken $K$-band spectroscopy of a sample of 13 pre-CVs in order to 
examine them for anomalous chemical abundances. In particular, 
we study the strength of the $^{12}$CO and $^{13}$CO absorption bands that 
have been found diminished and enhanced, respectively, in similar studies of 
CVs. 
}
{
All our systems show CO abundances that are within the range observed for
single stars. The weakest $^{12}$CO bands with respect to the spectral type
are found in the pre-CV BPM 71214, although on a much smaller scale than 
observed in CVs. Furthermore there is no evidence for enhanced $^{13}$CO.
Taking into account that our sample is subject to the present observational 
bias that favours the discovery of young pre-CVs with secondary stars of
late spectral types, we can conclude the following: 1) our study provides 
observational proof that the CO anomalies discovered in certain CVs are not 
due to any material acquired during the common envelope phase, and 
2) if the CO anomalies in certain CVs are not due to accretion of 
processed material during nova outburst, then the progenitors of these CVs
are of a significantly different type than the currently known sample of
pre-CVs.
}
{}


\maketitle

\section{Introduction}

A cataclysmic variable (CV) is a close interacting binary consisting
of a white dwarf (WD) and a red (K--M) main-sequence star (RD). The latter,
secondary star, transfers mass via Roche-lobe overflow to the more
massive white dwarf, the primary component. Typical orbital periods range 
from $\sim$80 min to $\sim$10 h. For comprehensive overviews on CVs
see \citet{warner95-1} and \citet{hellier01-1}.

The progenitors of CVs are thought to be initially wide, detached binaries. As 
the more massive component expands in the course of its nuclear evolution, it
fills its Roche lobe and transfers matter at high rates onto the red dwarf.
At mass-loss rates $\dot{M}_1 \sim 0.1~M_\odot~\mathrm{yr}^{-1}$, the
corresponding dynamical time scale is much shorter than the Kelvin-Helmholtz
time scale for thermal adjustment of the secondary star. A common-envelope (CE)
configuration results, providing an enhanced braking mechanism that rapidly
shrinks the orbital separation, until, at $P_\mathrm{orb} \la 1~\mathrm{d}$, 
the kinetic energy stored in the CE exceeds the binding energy,
and the CE is expelled as a planetary nebula, leaving a close, but still 
detached, WD/RD binary. The remaining mechanisms for angular-momentum loss, 
magnetic braking and/or gravitational radiation, further shrink the orbital 
separation, now on a much longer time scale, until the Roche lobe of the 
secondary star comes into contact with the stellar surface, initiating 
mass-transfer and the proper CV phase. Systems that have gone through such
a CE configuration and eventually will evolve into a CV, are known as
post-CE binaries. Using the criterion by \citet{schreiber+gaensicke03-1},
such objects are called pre-CVs if they evolve into CVs in less than
Hubble time ($\sim$13 Gyr) and can thus be regarded as representative
progenitors of the current CV population. 

While it had been originally assumed that the secondaries enter the CV phase
as main-sequence stars, during the past decade substantial evidence has
been amounted that a large fraction of the secondary stars in
long-period ($P_\mathrm{orb} > 3~\mathrm{h}$) CVs shows signs of nuclear 
evolution 
\citep{beuermannetal98-1,harrisonetal04-1,harrisonetal05-1}.
This would mean that the binary spends a much longer time in its pre-CV state
than hitherto assumed, and could provide the solution
to certain discrepancies between modelled and observed CV population
\citep[e.g.][]{patterson98-1,gaensicke05-1}.

In particular, the work by Harrison et al.\ revealed diminished $^{12}$CO
and -- in comparison -- enhanced $^{13}$CO absorption bands in the $K$ spectra
of dwarf novae with orbital periods $>3$ h, which they interprete as being
due to CNO processed material finding its way into the stellar photosphere
of the secondary star. Currently discussed scenarios that could lead to this
phenomenon can be divided in two principal categories: 1) nuclear evolution
of the secondary star, and 2) accretion of nuclear processed material.

The former implies that sufficient time for nuclear evolution has to pass
from the CE to the CV phase, and is only feasible for secondary stars of 
comparatively early spectral type ($\la$K0V) since dwarfs of later spectral 
type will not evolve within Hubble time \citep[e.g.][]{polsetal98-1}. As a 
consequence, the secondary star temporarily might become the more massive 
component after the CE phase, and go through a phase of thermal-timescale mass 
transfer \citep[][ the latter discuss this in the context of
anomalous abundances]{schenker+king02-1,schenkeretal02-1,gaensickeetal03-1,
harrisonetal05-2}. 

For the accretion scenario there are two principle sources of processed 
material: either the secondary star swept up processed material 
during the CE phase, or it accreted ejected material during nova eruptions.
These possibilities have been discussed by \citet{marks+sarna98-1}, who find 
that the significance of such effect strongly depends on the (unknown)
efficiency of the secondary to accrete such material. Furthermore, in the 
case of accretion from the CE, such material will be stripped away from the
secondary already in the early stages of the semi-detached CV phase
\citep[see also the respective discussion in][]{harrisonetal04-1}. 

Both the evolution scenario and accretion from the CE would also lead 
to anomalous chemical abundances in the progenitors of CVs, i.e.\ in 
pre-CVs. We here present $K$ band spectroscopy of a sample of pre-CVs 
to investigate the strength of the CO features in these systems.

\section{\label{samp_sec}The sample}

\begin{table*}
\caption[]{Previously known properties of the sample stars. Coordinates
(J2000.0) were taken from SIMBAD, $JHK_s$ magnitudes
are from the 2MASS database. Typical photometric errors are $\sim$0.03 mag
for $K_s$, and $\sim$0.036 for the colours. Uncertain orbital periods are 
marked with a colon.}
\label{samp_tab}
\setlength{\tabcolsep}{0.8ex}
\begin{tabular}{lllllllll}
\hline\noalign{\smallskip}
name & R.A. & DEC & $K_s$ & $J\!-\!H$ & $H\!-\!K_s$ & $P_\mathrm{orb}$ [h] 
& spType & References \\
\hline\noalign{\smallskip}
\object{BPM 6502}
            & 10 44 11 & $-$69 18 20 & 10.56 & 0.527 & 0.335 & 8.08  
&           & \citet{kawkaetal00-1}\\
\object{BPM 71214}
           & 03 32 43 & $-$08 55 40 &  9.30 & 0.655 & 0.233 & 4.33  
& M2.5V     & \citet{kawkaetal02-1}\\
\object{CC Cet}
              & 03 10 55 & $+$09 49 27 & 12.93 & 0.540 & 0.249 & 6.82  
& M4.5--M5V & \citet{safferetal93-1}\\
\object{EC 12477-1738}
       & 12 50 22 & $-$17 54 46 & 12.60 & 0.639 & 0.262 & 13.7: 
&           & \citet{tappertetal04-1}\\
\object{EC 13349-3237}
       & 13 37 51 & $-$32 52 22 & 13.25 & 0.669 & 0.129 & 11.4: 
&           & \citet{tappertetal04-1}\\
\object{EC 13471-1258}
       & 13 49 52 & $-$13 13 38 & 9.98  & 0.558 & 0.288 & 3.62  
& M3.5--M4V & \citet{kilkennyetal97-1}, \\
 & & & & & & & & \citet{odonoghueetal03-1}\\
\object{EC 14329-1625}
       & 14 35 46 & $-$16 38 17 & 10.87 & 0.580 & 0.288 & 8.4:  
&           & \citet{tappertetal06-2} \\
\object{LTT 560}
             & 00 59 29 & $-$26 31 01 & 11.86 & 0.521 & 0.270 & 3.54  
& M5.5V     & \citet{tappertetal06-1}, \citet{hoard+wachter98-1}\\
\object{NN Ser}
              & 15 52 56 & $+$12 54 44 & 16.17 & 0.653 & 0.086 & 3.12  
& M4.75V    & \citet{haefner89-1}, \citet{haefneretal04-1}\\
\object{P83l-57}
             & 03 34 34 & $-$64 00 56 & 11.54 & 0.594 & 0.204 &      
&           & \\
\object{RE 1016-053}
         & 10 16 29 & $-$05 20 27 & 9.77  & 0.617 & 0.220 & 18.9  
& M1.5V     & \citet{thorstensenetal96-1}\\
\object{RR Cae}
              & 04 21 06 & $-$48 39 08 & 9.85  & 0.572 & 0.296 & 7.29  
& $\ga$M6V  & \citet{bruch+diaz98-1}, \citet{bruch99-1}\\
 & & & & & & & M4V & \citet{maxtedetal07-1}\\
\object{UZ Sex}
              & 10 28 35 & $+$00 00 29 & 10.94 & 0.532 & 0.276 & 14.3  
& M4V       & \citet{safferetal93-1}\\
\hline\noalign{\smallskip}
\object{1E 2310.4-4949}
      & 23 13 17 & $-$49 33 16 & 8.92  & 0.623 & 0.219 & --    
& M3Ve      & \citet{stockeetal91-1}\\
\object{J223315.83-603224.0}
 & 22 33 16 & $-$60 32 24 & 10.74 & 0.660 & 0.155 & --    
& M2V       & \citet{oliveretal02-1}\\
\object{LP 759-25}
           & 22 05 36 & $-$11 04 29 & 10.72 & 0.607 & 0.329 & --    
& M5.5V     & \citet{kirkpatricketal95-1}\\
\hline\noalign{\smallskip}
\end{tabular}
\end{table*}

We have used the TPP catalogue \citep{kubeetal02-1} to search for 
confirmed and candidate pre-CVs that are observable from the southern
hemisphere. We have restricted our sample to confirmed pre-CVs with known 
orbital period, and excluded systems in nebulae and with primary components
other than white dwarfs. There are a number of exceptions to the first 
criterion, in that we include three systems with uncertain orbital periods, 
and one, as yet unconfirmed pre-CV candidate P83l-57 
\citep[also known as Ret1;][]{downesetal05-1}. These objects have been 
part of a project that aimed at confirming the pre-CV nature of a number of 
candidates, and finding the orbital period by photometric means 
\citep{tappertetal04-1,tappertetal06-2}. 
The light curves of EC 12477-1738, EC 13349-3237, and EC 14329-1625, showed 
the periodic modulations that are typical for the sinusoidal or ellipsoidal
variations in pre-CVs, although due to insufficient data no conclusive period 
could be determined. Initial observations of P83l-57 showed variations that
could be interpreted as part of a pre-CV light curve: a decline of $\sim$0.005 
mag over $\sim$5 h in one night, and a rise of similar dimensions in the 
subsequent night, and thus the object was included in our target list for the 
$K$ band spectroscopy. However, later observations could not confirm this 
variation, so that the pre-CV status of P83l-57 remains doubtful at the 
moment and needs to be clarified by future observations.

Previous studies have already provided an estimate of the spectral type of the
secondary star for about two thirds of the systems in our sample. All of them
are M dwarfs that have time scales for nuclear evolution 
$> t_\mathrm{Hubble} \sim 13~\mathrm{Gyr}$ \citep[e.g.][]{polsetal98-1}. 
Furthermore, most of these systems are relatively young objects, with
white dwarf cooling times of less than a few $10^8$ yr (except RR Cae and
LTT560, which are $\sim$1 Gyr old). Given that the typical time to
evolve into a semidetached CV configuration is several Gyr 
\citep[assuming the standard prescription for orbital angular momentum loss,]%
[]{schreiber+gaensicke03-1} , most of the systems have lived only
through a relatively small fraction of their pre-CV live.
In fact, only 
EC 13471-1258 and potentially BPM 71214 (depending on the model for angular 
momentum loss) have already passed more than half of their time as a post-CE 
binary. In this, our sample reflects the present observational bias towards 
systems with hot white dwarf primaries and/or comparatively late-type 
secondary stars \citep{schreiber+gaensicke03-1}. 
Our targets therefore do not represent the progenitors of CVs with 
anomalous abundances if scenario 1 (evolution) applies. A positive detection
of anomalous CO strengths in our targets would be a strong indication that
such material has been acquired by the secondary star during the CE phase
\citep{marks+sarna98-1}.

For comparison we observed three late-type M dwarfs with spectral 
types similar to those of our program objects. Table 
\ref{samp_tab} presents selected known properties of our targets.

\section{Observations and data reduction}

\begin{table*}
\caption[]{Log of observations, containing the date of the observations (start 
of night), the number of individual spectra, the exposure time for a single 
exposure, and the total exposure time. The last three columns give the 
corresponding atmospheric standard star, its spectral type, and its adopted 
effective temperature.}
\label{obs_tab}
\begin{tabular}{llllllll}
\hline\noalign{\smallskip}
object & date &  $n_{\rm data}$ & $t_\mathrm{exp}$ [s] & $t_\mathrm{tot}$ [s] 
& std & spType & $T_\mathrm{eff}$ [K] \\
\hline\noalign{\smallskip}
BPM 6502            & 2005-12-21 & 2  & 10  & 20
& \object{Hip030743} & B4V     & 17\,000 \\
                    & 2006-01-12 & 2  & 10  & 20
& \object{Hip031068} & B3V     & 19\,000 \\
BPM 71214           & 2005-11-17 & 2  & 5   & 10
& \object{Hip026939} & B5V     & 15\,200 \\
CC Cet              & 2005-10-13 & 10 & 60  & 600
& \object{Hip024809} & B9V     & 10\,300 \\
                    & 2005-11-12 & 10 & 60  & 600
& \object{Hip034669} & B4V     & 17\,000 \\
EC 12477-1738       & 2005-03-28 & 10 & 60  & 600 
& \object{Hip065475} & B2IVn   & 21\,000 \\
EC 13349-3237       & 2005-04-18 & 10 & 60  & 600
& \object{Hip055051} & B1V     & 25\,500 \\
EC 13471-1258       & 2005-04-18 & 2  & 5   & 10
& Hip055051 & B1V     & 25\,500 \\
EC 14329-1625       & 2005-04-18 & 2  & 10  & 20
& Hip055051 & B1V     & 25\,500 \\
LTT 560             & 2005-06-01 & 2  & 30  & 60
& \object{Hip104320} & B3V     & 19\,000 \\
NN Ser              & 2005-03-28 & 24 & 300 & 7200
& \object{Hip081362} & B0.5III & 27\,750 \\
P83l-57             & 2005-11-17 & 2  & 30  & 60
& Hip026939 & B5V     & 15\,200 \\
                    & 2005-11-22 & 2  & 30  & 60
& \object{Hip015188} & B3V     & 19\,000 \\
RE 1016-053         & 2005-12-24 & 2  & 5   & 10
& \object{Hip050780} & B3V     & 19\,000 \\
                    & 2006-01-12 & 2  & 5   & 10
& \object{Hip033575} & B2V     & 21\,000 \\
RR Cae              & 2005-11-17 & 2  & 5   & 10
& Hip026939 & B5V     & 15\,200 \\
UZ Sex              & 2006-01-12 & 2  & 10  & 20
& Hip033575 & B2V     & 21\,000 \\
\hline\noalign{\smallskip}
1E 2310.4-4949      & 2005-05-26 & 2  & 5   & 10
& \object{Hip088426} & G0V     & 5\,940 \\
J223315.83-603224.0 & 2005-05-25 & 2  & 10  & 20
& \object{Hip095103} & G3V     & 5\,700 \\
LP 759-25           & 2005-05-25 & 2  & 10  & 20
& \object{Hip105633} & B2/B3V  & 20\,200 \\
\hline\noalign{\smallskip}
\end{tabular}
\end{table*}

The data were obtained with ISAAC mounted at Antu (UT1), VLT, Paranal, Chile.
The instrument was operated in SWS (short-wavelength spectroscopy) mode,
and the grating was used in its low-resolution
(resolving power $\sim$1500),
$K$-band, position. The
nominal wavelength coverage was $\sim$1.85--2.57 $\mu$, though only data in
the wavelength range  $\sim$2.00--2.45 $\mu$ were useful. 
Observations were 
conducted in service mode and included flat fields and wavelength 
calibration (Xe-Ar) data at the start of night, and telluric standard 
stars taken within 1 h and an airmass difference $\Delta M(z) = 0.2$ of 
the target spectra. The data were taken in AB--BA cycles, i.e.\, with small 
telescope offsets after the first and then after every second spectrum, so 
that spectra 1, 4, 5, 8, ..., occupy positions in the lower half of the CCD 
(position A), and spectra 2, 3, 6, 7, ..., are located in the upper half of 
the CCD (position B). Some stars were observed twice, since the first 
observations did not match ESO's quality criteria (regarding, e.g., seeing, 
difference in airmass between target and telluric standard, etc.). In one 
case (CC Cet), both spectra were found to be of sufficient quality, and could 
be combined in order to improve the S/N. For another system (BPM 6502) the 
spectra of the telluric standards presented significant disturbances on both 
occasions, fortunately once affecting mostly the blue part, and the other 
time mostly the red part of the spectrum. 
The latter spectrum was used for the SED analysis, since the spectral slope
remained mostly intact. On the other hand, in the former spectrum, the 
spectral lines were not affected, and so this spectrum was used to measure
equivalent widths.
In the the cases of P831-57 and RE1016-053, only the second epoch data 
were useful. Table \ref{obs_tab} presents a summary of the observations.

The reduction was done with IRAF packages. After flatfielding, a 
two-dimensional wavelength calibration was applied to the object data, in 
order to correct for the positional dependence of the dispersion. 
The resulting, "straightened", AB pairs of a specific object were then
subtracted from each other and, after corresponding offsets had been
applied, combined to a single image. Subsequently, the spectra of the
targets and the telluric standards were extracted. For some days, no
Xe-Ar wavelength calibration was provided. In these cases, it was found that
calibration data from other days were sufficient to perform the 2-D 
transformation, but that it was necessary to apply a zero point correction
to the extracted data using the night sky lines \citep{rousselotetal00-1}.

With Br$\gamma$ at 2.17 $\mu$, the telluric standards of spectral type B have 
basically only one intrinsic absorption line in the $K$ band. The very 
early type B stars also show a very weak He{\sc I} line at 2.11 $\mu$. In
both cases, those lines were fitted with a Voigt profile and subsequently
subtracted from the spectrum. For the standards of spectral type G, a solar
spectrum (NSO/Kitt Peak FTS data) was rebinned and smoothed down to the 
resolution of the 
ISAAC spectra, shifted in wavelength to correct for different zero points, and
finally subtracted from the telluric spectra. The resulting pure atmospheric
absorption spectra then were shifted and scaled to match position and depth
of the atmospheric features in the corresponding target spectrum. Reference
points for the shifting were the narrow absorption lines in the red
part of the spectrum, while the broad feature between 2.0 and 2.1 $\mu$
was used to adjust for the depth of the atmospheric absorption. Finally,
the target spectra were divided by the telluric spectra, and, in order to 
recover the intrinsic SED of the targets, multiplied
with a blackbody spectrum corresponding to the effective temperature of the
telluric standard (see Table \ref{obs_tab}).

\section{Results}

\subsection{Spectral types}

\begin{table}
\caption[]{Estimated spectral type and corresponding effective temperature
of the targets based on their SED. For comparison, the spectral types from the
literature (listed in Table \ref{samp_tab}) are repeated here (in brackets). 
The error in the temperature column corresponds to the estimated range. The 
last column gives the equivalent width (in {\AA}) of the 
Na{\sc I} $\lambda$2.21 $\mu$ absorption line. The table is sorted with 
respect to the strength of the latter.}
\label{eqw1_tab}
\setlength{\tabcolsep}{0.8ex}
\begin{tabular}{llll}
\hline\noalign{\smallskip}
object & spType & $\log T_\mathrm{eff}$ & Na{\sc I} \\
\hline\noalign{\smallskip}
RR Cae              & M3--M4.5 ($\ga$M6V)    & 3.510(10) & 3.4  \\
UZ Sex              & M2.5--M5 (M4V)         & 3.515(25) & 3.5  \\
EC 13349-3237       & K2--M1                 & 3.590(60) & 3.9  \\
RE 1016-053         & K1--K5 (M1.5V)         & 3.670(30) & 4.1  \\
BPM 71214           & K2--M1 (M2.5V)         & 3.590(60) & 4.9  \\
LTT 560             & M5.5--M6 (M5.5V)       & 3.430(20) & 4.9  \\
EC 13471-1258       & M3.5--M5 (M3.5--M4V)   & 3.485(25) & 5.2  \\
BPM 6502            & M2.5--M5               & 3.500(40) & 5.6  \\
EC 12477-1738       & M3.5--M5               & 3.475(15) & 6.0  \\ 
P83l-57             & M2.5--M3.5             & 3.510(30) & 6.4  \\
CC Cet              & M3.5--M5.5 (M4.5--M5V) & 3.480(30) & 7.8  \\
EC 14329-1625       & M3.5--M4.5             & 3.495(15) & 8.1  \\
\hline\noalign{\smallskip}
LP 759-25           & M5.5--M6 (M5.5V)       & 3.430(20) & 4.3  \\
J223315.83-603224.0 & M1--M2.5 (M2V)         & 3.530(10) & 4.6  \\
1E 2310.4-4949      & M3--M4.5 (M3Ve)        & 3.485(25) & 5.2  \\
\hline\noalign{\smallskip}
\end{tabular}
\end{table}

\begin{figure}
\includegraphics[angle=-90,width=\columnwidth]{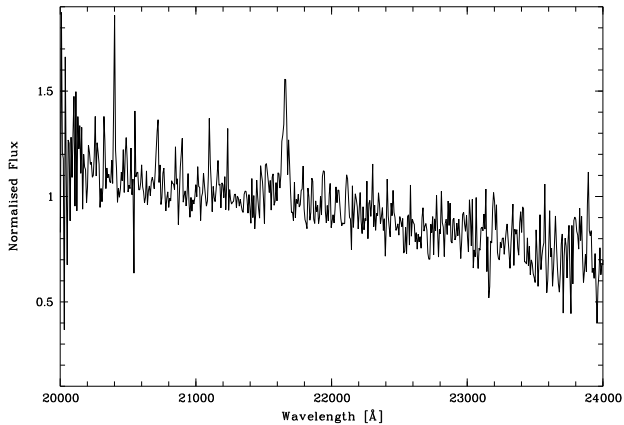}
\caption[]{Unsmoothed $K$-band spectrum of NN Ser. The only detected
spectral feature is the Br$\gamma$ emission line.}
\label{nnser_fig}
\end{figure}

\begin{figure*}
\includegraphics[angle=-90,width=2.0\columnwidth]{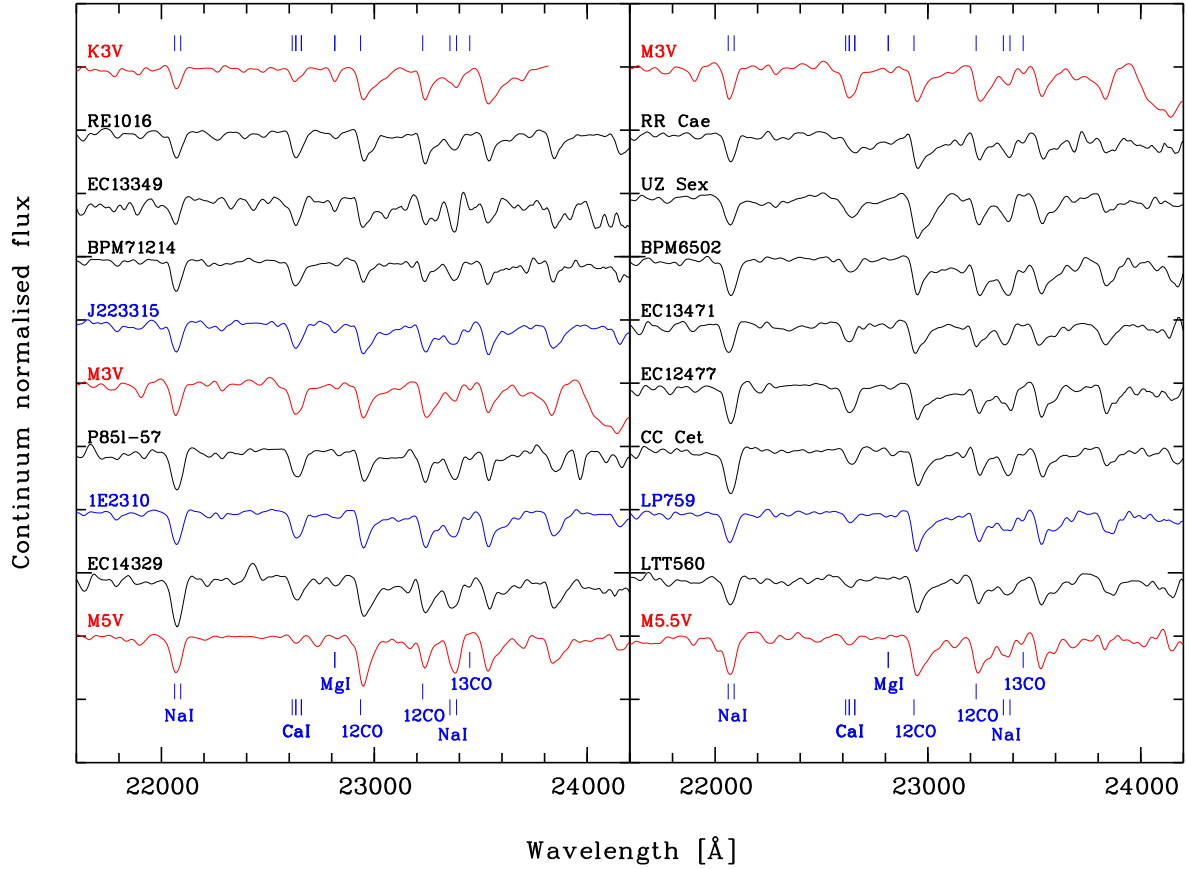}
\caption[]{Continuum-normalised target spectra. The data have been smoothed
to match the resolution of \citet{ivanovetal04-1}. Spectra are roughly
sorted according to their estimated spectral type. For comparison, the plot
also includes the following objects from Ivanov et al.: \object{HR8832} (K3V),
\object{GJ388} (M3V), \object{GJ866} (M5V), and \object{GJ905} (M5.5V).
}
\label{spec_fig}
\end{figure*}

\begin{figure}
\includegraphics[angle=-90,width=\columnwidth]{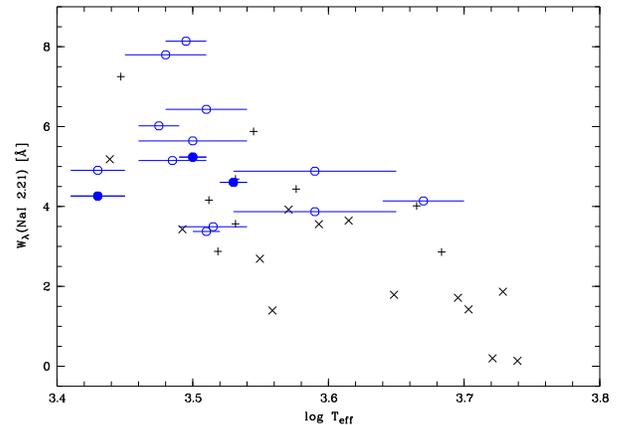}
\caption[]{Equivalent width of Na{\sc I} as a function of effective 
temperature. Stars from \citet{ivanovetal04-1} with $-0.1 \le \mathrm{[Fe/H]} 
\le 0.1$ are marked by $+$, those with metallicities outside this range, or, 
in the one case of the star with the lowest $T_\mathrm{eff}$, unknown,
with $\times$. Open circles indicate the pre-CVs from our sample, filled ones 
represent the three comparison late-type dwarfs.}
\label{NaI_fig}
\end{figure}

\begin{figure}
\includegraphics[angle=-90,width=\columnwidth]{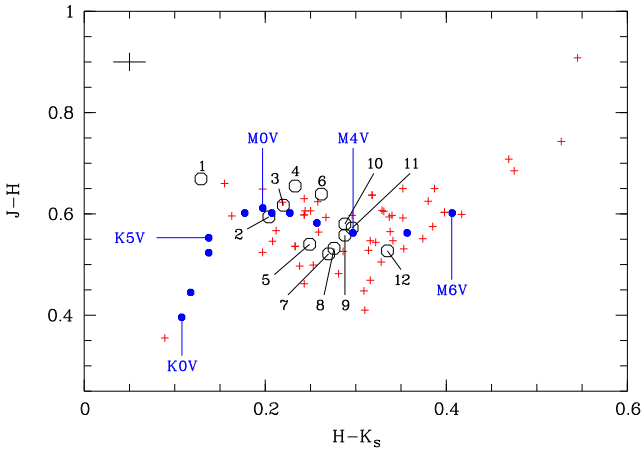}
\caption[]{Colour-colour diagram of the 2MASS $JHK_s$ photometry. The plot 
includes late-type dwarfs from spectroscopic libraries (crosses), 
main-sequence colours (filled circles), and pre-CVs (open circles). The 
latter are labelled in sequence of increasing $H\!-\!K_s$ as follows: 
(1) EC 13349-3237, (2) P83l-57, (3) RE 1016-053, (4) BPM 71214,
(5) CC Cet, (6) EC 12477-1738, (7) LTT 560, (8) UZ Sex, (9) EC 13471-1258
(10) EC 14329-1625, (11) RR Cae, (12) BPM 6502. Only systems with 
photometric errors $<$0.05 in either of the colours are shown. The cross
in the upper left corner indicates the average error $\sim$0.036 mag for
the objects included in this plot.
}
\label{jhk_fig}
\end{figure}

\begin{table}
\caption[]{Comparison of the spectral types estimated via the 2MASS 
colour-colour diagram, the $K$ band spectral energy distribution, and the
line strengths in the $K$ spectra. The last columns gives previous estimates
from the literature (for references see Table \ref{samp_tab}). Systems that
lie above or below the main-sequence in the two-colour diagram are marked
by "$+H$" and "+blue", respectively.}
\label{jhk_tab}
\setlength{\tabcolsep}{0.8ex}
\begin{tabular}{lllll}
\hline\noalign{\smallskip}
object & $JHK$ & SED & lines & literature \\
\hline\noalign{\smallskip}
BPM 6502      & M4--M5 +blue & M2.5--M5   & M5       & --        \\
BPM 71214     & M2 +$H$      & K2--M1     & M3       & M2.5     \\
CC Cet        & M3.5 +blue   & M3.5--M5.5 & M3--M5.5 & M4.5--M5 \\
EC 12477      & M3 +$H$      & M3.5--M5   & M3       & --        \\ 
EC 13349      & $>$K5 +$H$   & K2--M1     & K5--M2   & --        \\
EC 13471      & M4           & M3.5--M5   & M2       & M3.5--M4 \\
EC 14329      & M4           & M3.5--M4.5 & M3       & --        \\
LTT 560       & M4 +blue     & M5.5--M6   & M5.5     & M5.5     \\
P83l-57       & M1           & M2.5--M3.5 & M3       & --        \\
RE 1016       & M1.5         & K1--K5     & K5--M2   & M1.5     \\
RR Cae        & M4           & M3--M4.5   & M4       & $\ga$M6 / M4  \\
UZ Sex        & M4 +blue     & M2.5--M5   & M4       & M4       \\
\hline\noalign{\smallskip}
\end{tabular}
\end{table}

Based on optical spectra, earlier studies have provided estimates of the 
spectral type for the majority of the targets in our sample. 
To obtain independent estimates for the spectral types of the secondary
stars in our program pre-CVs, we have compared our $K$ spectra
to the spectral catalogues
of \citet{leggettetal00-1}, 
\citet{kleinmann+hall86-1}, and \citet{ivanovetal04-1} \footnote{See 
http://ftp.jach.hawaii.edu/ukirt/skl/dM.spectra/ for Leggett's data, the 
other two are available via CDS.} (hereafter L00, KH86, and I04, respectively).
Each catalogue has strengths and weaknesses for this application. The L00 data 
represent the best coverage of spectral subtypes, but are limited to M dwarfs,
and have very low spectral resolution. The I04 catalogue still provides a very 
acceptable number of K and M dwarfs, at an only slightly lower spectral 
resolution than our data. However, their spectra are normalised with respect 
to the continuum slope, and thus there is no information on the SED. Finally, 
the KH86 sample contains only 4 late-type dwarfs, but provides the highest 
spectral resolution, and, although the library data are continuum normalised, 
the SED can be recovered by multiplying the spectra with the blackbody 
spectrum of an A0V star \citep{foersterschreiber00-1}. We therefore estimated 
the spectral type (and $T_\mathrm{eff}$) of our targets by comparing their 
spectral energy distribution to the L00 and KH86 dwarfs, and tested this 
estimate using the equivalent width of the Na{\sc I} $\lambda$2.21$\mu$ 
absorption line in the I04 spectra.

For the comparison of the SED, we first shifted our data to the rest
wavelength of the Na{\sc I} $\lambda$2.21$\mu$ line, then smoothed our
and KH86's data to match the resolution of the L00 data, and finally 
normalised all three data sets by dividing through the average
flux value of the 2.10--2.14 $\mu$ wavelength interval.
The results of the visual comparison are summarised in Table \ref{eqw1_tab}.
This, and the subsequent analysis, does not
include the object NN Ser, since the S/N proved too low for the detection
of absorption features. For completeness, we present its unsmoothed spectrum
in Fig.\ \ref{nnser_fig}.

Such visual comparison over a limited spectral range can certainly yield
only a very rough estimate of the spectral type. Since several members of
our sample have been found to show significant irradiation by the white dwarf,
one should furthermore expect that those stars appear somewhat bluer, and
that the corresponding temperatures will be overestimated.

We can test these estimates by measuring the strength of suitable absorption
features in our spectra. In the $K$ band, the Na{\sc I} $\lambda$2.21$\mu$
line appears as the best choice, since it shows a distinctive dependence 
of temperature, but is independent of luminosity class, and thus
nuclear evolution \citep[ their Fig.9]{ivanovetal04-1}. The stars in the
I04 library were taken with the same instrumentation as our targets, although
at a slightly lower spectral resolution, and we smoothed our spectra
correspondingly. We then normalised the spectra for their SED by fitting
splines to the continuum and dividing by the fit. These spectra are shown
in Fig.\ \ref{spec_fig}. Equivalent widths were measured using the index 
definition from \citet{alietal95-1} as listed in I04. The results
are summarised in Table \ref{eqw1_tab} and plotted in Fig.\ \ref{NaI_fig}, 
together with the stars from the I04 catalogue.

Although this index presents a large scatter even within the library
stars, the plot does show that the pre-CVs on the average appear to have
slightly higher equivalent widths at a given temperature.
With CC Cet and EC 14329-1625 there are two systems with
exceptionally strong Na{\sc I} absorption, be it due to enhanced Na{\sc I},
or due to a much later spectral type than estimated 
\citep[note, however, that our estimate for CC Cet fits well the
result from][]{safferetal93-1}. On the other hand the two confirmed M5.5-M6 
dwarfs LTT 560 and LP 759-25 (the latter being one
of the comparison stars) have a comparatively shallow Na{\sc I} absorption
line. Still, on the whole, our estimates appear consistent with the 
behaviour of the Na{\sc I} spectral index.

The referee suggested to use the 2MASS $JHK$ database in order to further 
explore the possibility that irradiation by the primary alters the intrinsic
SED in the $K$ band, thus causing an overestimation of the temperature 
in our targets. In Fig.\ \ref{jhk_fig}, we present the corresponding
colour-colour diagram to compare our targets with the late-type dwarfs from
the spectroscopic catalogues of \citet{alietal95-1}, 
\citet{kleinmann+hall86-1}, \citet{ivanovetal04-1}, and 
\citet{leggettetal00-1}. Following \citet{hoardetal02-1}, we have also
included the main-sequence from \citet[ p.151]{cox00-1}, converted to
the 2MASS photometric system using the transformations from
\citet{carpenter01-1}. Irradiation would make the secondary stars in pre-CVs
appear bluer, and thus result into a displacement towards the lower left
of the main sequence. We do find four of
our targets in this direction, but still well within the general scatter
that is observed also for single late-type stars. Three targets lie
somewhat above the main-sequence, i.e.\ they show an excess in their
$H$ magnitude, the most extreme case being EC 13349-3237. 
A second system worth mentioning is LTT 560. As we discuss in
Section \ref{co_sec}, its $K$ band spectrum is very similar to the
M5.5 dwarf LP 759-25. The near-infrared colours of the two stars, however,
do not match, with LTT 560 being distinctively bluer (Table \ref{samp_tab}),
and it appears in Fig.\ \ref{jhk_fig} as a blueshifted M4 dwarf.
Still, this system contains a very cool white-dwarf primary
\citep[$T_\mathrm{WD} \sim 7500$ K][]{tappertetal07-1}, and this 
displacement can therefore not be due to irradiation. Note also that
the primary of RR Cae has a very similar temperature, and since this object
does not appear blueshifted, it is unlikely that contribution from the
white dwarf itself is causing this shift in LTT 560. Photometric
light curves show evidence for flaring activity, and so the displacement
might be explained by 2MASS having caught LTT 560 during an active phase.

In Table \ref{jhk_tab} we compare the spectral types of our targets
determined with the three different methods. If irradiation had any
effect on the $K$-band SED, we would expect that the spectral-type estimates
from 2MASS and from SED agree well with each other, but not with the 
estimates from the line strengths. The fact, that we find all three
methods providing very similar results ($\pm$1 subclass) for most of the 
systems shows that this is not the case.

\subsection{The $^{12}$CO absorption\label{co_sec}}

\begin{figure*}
\includegraphics[width=1.80\columnwidth]{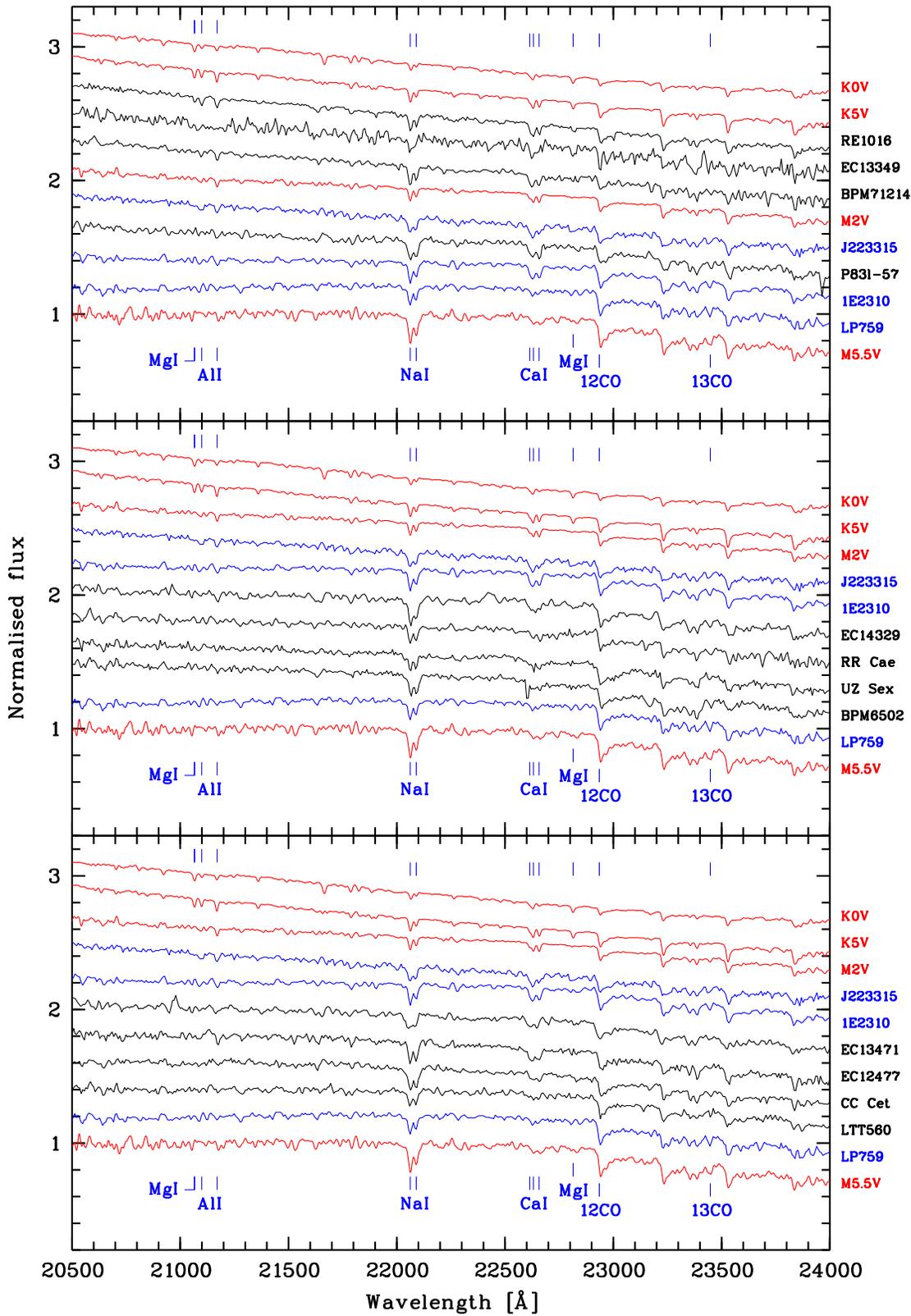}
\caption[]{Unsmoothed, normalised spectra. Each plot includes four pre-CVs,
the three comparison stars, and the four late-type dwarfs from the
\citet{kleinmann+hall86-1} catalogue. The individual spectra are vertically 
displaced by 0.2 units. The sequence roughly corresponds to the estimated 
spectral type, with the plot at the top containing the earliest, and the one 
at the bottom the latest. 
Note that the M2V standard, \object{Gl 411}, has a rather low metallicity 
of $-$0.33 
dex \citep{bonfilsetal05-1}, resulting in generally weaker absorption
features.
}
\label{kh_fig}
\end{figure*}

The principal result of the $K$-band spectroscopy of cataclysmic variables
by \citet{harrisonetal04-1,harrisonetal05-2,harrisonetal05-1} was the unusual
weakness of the $^{12}$CO absorption together with enhanced $^{13}$CO.
While a more quantitative method would in principle be desirable, we here
follow the approach by Harrison et al.\ and visually compare our target
spectra to single stars of similar spectral type. The reason for this is that
the only available library that includes a large number of late-type dwarfs 
at sufficient spectral resolution by \citet{ivanovetal04-1} contains continuum 
normalised data. For the comparison of the Na{\sc I} absorption, this did not
pose a great difficulty, since the blue part can be fitted relatively easily.
Furthermore, the slope of the Na{\sc I} relation with temperature is steep,
making the Na{\sc I} strength a comparatively robust parameter. In contrast,
in the red part of the spectrum, the continuum is not well defined due to the 
extended and overlapping CO absorption bands. Systematic differences between 
the library stars and our data are thus easily introduced right in the 
spectral range that is of most interest.

We therefore turn to the aforementioned visual approach and in 
Fig.\ \ref{kh_fig} present a comparison of our unsmoothed spectra with the 
KH86 data. For this purpose, the latter have been smoothed to match our 
spectral resolution. In the following we summarise the results for each 
object in detail. 
\smallskip\\
{\em BPM 6502:} This object was observed twice, unfortunately both times
with unsatisfactory results. The first spectrum showed a strong deviation
of the SED in the blue part of the spectral range, while a broad 'emission'
feature affected the red part (2.27--2.35 $\mu$). We attempted to remove the
latter by fitting a broad Gaussian, and this spectrum is shown in Fig.\ 
\ref{kh_fig} (middle plot). There remained, however, notable differences in 
comparison with the -- apparently unaffected -- red part of the first 
spectrum, so the latter was used to measure equivalent widths (and is 
presented in Fig.\ \ref{spec_fig}). 

There is no previous estimate on the spectral type of the secondary in this 
system, but \citet{kawkaetal00-1} find a mass to $M_2 = 0.16(09) M_\odot$, 
indicative of an $\sim$M5-M6 dwarf. The $K$-band SED points to a somewhat 
earlier type. However, as explained above, that SED is not entirely 
trustworthy. Indeed, the Na{\sc I} and Ca{\sc I} line strengths suggest 
a spectral type close to M5V, since they are similar to the M5.5V standard
from KH86 (Na{\sc I} is a bit weaker, and Ca{\sc I} slightly stronger). All
CO absorption bands show normal strengths. 
\smallskip\\
{\em BPM 71214:} \citet{kawkaetal02-1} give M2.5V as spectral type. Again,
our SED analysis yields an earlier type, but the line strengths (Na{\sc I},
Ca{\sc I}) are very similar to the M3Ve star 1E2310.4-4949, 
favouring the Kawka et al.\, result. This is supported by the
weakness of the Mg{\sc I} $\lambda$2.11/2.28 $\mu$ lines, which are barely,
if at all, detected. On the other hand, the CO features are very weak for such 
spectral type and fit much better the K5V star from KH86. 
\smallskip\\
{\em CC Cet:} This object was also observed twice. Both spectra were of
sufficient quality, so that they could be combined in order to increase the
S/N. The spectral type suggested by the SED agrees well with the
previous estimate of M4.5--M5 by \citet{safferetal93-1}. Also the line
strengths place the object between the M3Ve star 1E2310.4-4949 and KH86's
M5.5 dwarf. In comparison, the CO absorption appears slightly too weak.
\smallskip\\
{\em EC 12477-1738:} The spectroscopic characteristics are
similar to CC Cet. The stronger Ca{\sc I} indicates a slightly earlier type,
probably closer to M3V than M5.5V. CO appears at about the same strength as
in CC Cet. 
\smallskip\\
{\em EC 13349-3237:} The faintest member of our sample (apart from NN Ser),
and this is unfortunately reflected in the low S/N. SED and line strengths
both place it somewhere between the KH86's K5V and M2V, with the clearly
detected Mg{\sc I} $\lambda$2.28 $\mu$ line pointing to the earlier limit
of this range. Worth noting is furthermore the non-detection of the Al{\sc I}
$\lambda$2.11/2.12 $\mu$ lines. These have lesser strength than Mg{\sc I} only
for spectral types earlier than K5V. In contrast, CO bands are clearly visible,
although the low S/N impedes a more precise comparison of their strength.
\smallskip\\
{\em EC 13471-1258:} \citet{odonoghueetal03-1} found that this system is 
already close to start its CV phase, and estimated its spectral type to
M3.5--M4V. The absorption features in our spectrum, comparatively weak 
Na{\sc I}, Ca{\sc I}, and CO, place it close to the M2V star 
J223315.83-603224.0, although 
both the 2MASS colours and the $K$-band SED agree better with the former 
estimate. 
\smallskip\\
{\em EC 14329-1625:} The spectrum shows great similarities with
the M3Ve star 1E2310.4-4949, with the notable exception of the enhanced
Na{\sc I} line.
\smallskip\\
{\em LTT 560:} This object is almost a spectroscopic twin to the M5.5 dwarf 
LP 759-25, with only slightly weaker absorption lines and bands. 
LTT 560 is a remarkable system in many aspects: it contains the coolest
white dwarf in a pre-CV besides RR Cae, and there is evidence for stellar
activity and low-level mass transfer, although the secondary star does not
fill its Roche lobe \citep{tappertetal07-1}. Its $K$-band spectrum,
however, does not show any anomalies.
\smallskip\\
{\em P83l-57:} Na{\sc I} and Ca{\sc I} line strengths are similar to the
M3Ve star 1E2310.4-4949, while the CO bands resemble more those in the
M2 dwarf J223315.83-603224.0. Initial suspicions about photometric variability
in the form of a sinusoidal or ellipsoidal light curve \citep{tappertetal04-1}
could not be confirmed. Since there are a several narrow emission lines 
detected in the optical spectrum \citep[e.g., Ca H and K, and the Balmer 
series;][]{rodgers+roberts94-1}, this object is either a wide, detached, 
binary with a very active red dwarf, or -- somewhat more likely -- seen at
low orbital inclination. 
Note also that the 2MASS data indicates a slightly earlier type (M1V)
which could be due to irradiation, implying that this system indeed
has a hot companion.
\smallskip\\
{\em RE 1016-053:} Both SED and the Na{\sc I} and Ca{\sc I} are very similar
to BPM 71214, although the presence of the Mg{\sc I} lines indicates an
earlier type. Comparison with the KH86 stars on the basis of the Mg{\sc I}
strength with respect to Al{\sc I} places the star somewhat later than K5V 
and somewhat earlier than M2V, in good agreement with 
\citet{thorstensenetal96-1}, who found M1.5V. The CO bands appear at normal
strength, and stronger than in BPM 71214, emphasising their weakness in the
latter star.
\smallskip\\
{\em RR Cae:} The SED of this star fits best with the M3.5V standards from
the L01 library. \citet{bruch+diaz98-1} find $\ga$M6V, but this appears
unlikely, since the blue part of the spectrum does not show any evidence
of the H$_2$O depression that is typical for late M dwarfs. RR Cae contains 
a very cool white dwarf primary \citep[$T_\mathrm{WD} \sim 7000$ 
K;][]{bragagliaetal95-1}, so that there are no irradiation effects present
that could alter the intrinsic slope of the secondary's continuum. Both
SED and line strengths are similar to UZ Sex, which has been classified as
M4V \citep{safferetal93-1}. 
Furthermore, a recent study on optical spectra of RR Cae by 
\citet{maxtedetal07-1} also finds an M4V secondary star, in good agreement
with our infrared data. For such spectral type, the CO bands show normal
abundances.
\smallskip\\
{\em UZ Sex:} As mentioned above, this is probably an M4V star with perfectly
normal abundances.
\smallskip\\

We close this section with the remark that, while we detect $^{13}$CO in
all stars in our sample, none of the systems shows it at anomalous strength.

\section{\label{disc_sec}Discussion and conclusion}

With BPM 71214 we find one system in our sample that at first glance appears
as a promising candidate for diminished $^{12}$CO. There seem to be 
certain problems with the reduction for telluric features, as indicated by 
two unexplained absorption lines at 2.318 $\mu$ and 2.372 $\mu$ 
(Fig.\ \ref{kh_fig}). However, if additional telluric absorption should 
also affect the CO band, this would result in enhanced absorption, and not 
in the observed diminished one. 
In any case, this potential depletion of CO is not nearly as dramatic as found 
in certain CVs \citep{harrisonetal04-1,harrisonetal05-1}. Taking into account
the spread of CO line strengths in single late-type dwarfs 
\citep[ their Fig.\ 9]{ivanovetal04-1}, and also the fact that none of our 
systems shows any enhancement of $^{13}$CO with respect to $^{12}$CO, 
we conclude that, at least regarding the CO abundance, all pre-CVs in our 
sample are consistent with main-sequence stars.

A comparatively large fraction of our targets appears to have abnormally 
strong Na{\sc I} absorption (Fig.\ \ref{NaI_fig}, Table \ref{eqw1_tab}). 
While in three systems (RE 1016$-$053, BPM 71214, P83l$-$57) such potential 
enhancement is unclear due to the uncertainty regarding their spectral type,
both CC Cet and especially EC 14329$-$1625 inhabit a stronger Na{\sc I} line 
than any other star in the \citet{alietal95-1}, \citet{ivanovetal04-1}, and
\citet{kleinmann+hall86-1} catalogues. All these catalogues only include
M dwarfs up to $\sim$M5.5, but \citet{cushingetal05-1} have shown that
Na{\sc I} has maximum strength at $\sim$M6V and diminishes
towards later spectral types, disappearing completely around $\sim$L1. The
enhanced line in CC Cet and EC 14329$-$1625 is therefore not due to an
erroneous assignation of the spectral type. However, since the uncertainties 
in the spectral type for the three above mentioned systems are comparatively
large, and since the effect in CC Cet is not overly dramatic 
\citep[ give $W_\mathrm{\lambda,NaI} = 7.6 \pm 0.2~\mathrm{\AA}$ for the M6V 
star Gl 406, while CC Cet has $W_\mathrm{\lambda,NaI} = 7.8~\mathrm{\AA}$]%
{cushingetal05-1}, it is well possible that this apparent anomaly of a group
of pre-CVs melts down to just one peculiar object, EC 14329$-$1625.

In agreement with previous results, most pre-CV secondary stars in our sample 
turned out to have spectral types $\ga$M2V, and therefore will not evolve 
within Hubble time \citep[e.g.][]{polsetal98-1}. 
As discussed in Section \ref{samp_sec}, we therefore did not expect to be able 
to confirm scenario 1 (nuclear evolution of the secondary star). 
The possibility that processed material is accreted by the secondary star
during the CE phase has been investigated in detail by 
\citet{marks+sarna98-1}, who find that such potential accretion can not 
account for the abundance anomalies observed in CVs, since the accreted
material will be stripped from the secondary star during the initial
stages of mass-transfer. Our $K$-band spectra of pre-CVs now show that only a
very small amount of CE material, if any, is accreted by the secondary, since
it leaves no trace already in comparatively young pre-CVs. 

There remain therefore two possibilities for the presence of anomalous
CO strengths in certain CVs. Either these systems originate from a very
different type of pre-CV 
\citep[e.g., supersoft binaries;][]{schenkeretal02-1}, 
or the material was accreted during nova eruptions. 

Assuming for a moment the former, we point out that \citet{harrisonetal05-2} 
find that magnetic CVs have normal abundances, motivating them to suggest 
different formation mechanisms for magnetic and non-magnetic CVs. We here draw 
attention to the fact that there is no indication for any strong magnetic 
white dwarf primary in the systems of our sample \citep[and they appear 
to be pretty rare in pre-CVs in general;][]{liebertetal05-1, 
silvestrietal07-1}. 
These objects therefore will eventually evolve into non-magnetic CVs without 
anomalous abundances. This does not invalidate the argumentation by 
\citet{harrisonetal05-2}, but indicates that the evolution of CVs 
might even come in more flavours than hitherto suspected.

Our study furthermore emphasises the need for more data, both with respect
to single late-type dwarfs in order to better address the abundance scatter
within a specific spectral type, and regarding the large discrepancy between
the number of known pre-CVs and CVs \citep[e.g.,][]{morales-ruedaetal05-1,
ritter+kolb03-1}. Thanks to the Sloan Digital Sky Survey, the latter picture
is improving tremendously \citep{silvestrietal06-1,silvestrietal07-1}. 
However, even this much enlarged sample of pre-CVs is still strongly 
biased towards 'young' pre-CVs and late-type secondary stars 
\citep{schreiber+gaensicke03-1,schreiberetal06-1}, and further studies
will be necessary in order to establish a representative sample of 
CV progenitors.

\begin{acknowledgements}
We thank the anonymous referee for valuable comments that helped to improve
the paper. Many thanks to Valentin Ivanov for making an early version of 
his spectral library available, and to Tom Harrison for helpful discussion
under difficult conditions. Further thanks to Carsten Weidner for insight 
in evolutionary time scales.

CT and RM acknowledge financial support by FONDECYT grant 1051078.  BTG was
supported by a PPARC Advanced Fellowship.  
This work has made intensive use of the SIMBAD database, operated at CDS, 
Strasbourg, France, and of NASA's Astrophysics Data System Bibliographic 
Services. 
This publication makes use of data products from the Two Micron All Sky Survey,
which is a joint project of the University of Massachusetts and the Infrared 
Processing and Analysis Center/California Institute of Technology, funded by 
the National Aeronautics and Space Administration and the National Science 
Foundation.
IRAF is distributed by the National Optical Astronomy Observatories.
NSO/Kitt Peak FTS data used here were produced by NSF/NOAO.
\end{acknowledgements}


\end{document}